\documentclass[12pt]{iopart}
\usepackage{setstack,iopams}
\usepackage{mathrsfs}
\usepackage[dvips]{graphicx}

\newenvironment{ul}[1]{
\vspace{-0.5ex}
\begin{itemize}
\setlength{\partopsep}{#1ex}
\setlength{\topsep}{#1ex}
\setlength{\itemsep}{#1ex}
\setlength{\parsep}{#1ex}
\setlength{\parskip}{#1ex}}{
\end{itemize}
\vspace{-0.5ex}}

\newenvironment{ol}[1]{
\vspace{-0.5ex}
\begin{enumerate}
\setlength{\partopsep}{#1ex}
\setlength{\topsep}{#1ex}
\setlength{\itemsep}{#1ex}
\setlength{\parsep}{#1ex}
\setlength{\parskip}{#1ex}}{
\end{enumerate}
\vspace{-0.5ex}}

\newtheorem{define}{Definition}[section]
\newtheorem{theorem}[define]{Theorem}
\newtheorem{lemma}[define]{Lemma}

\newtheorem{corollary}[define]{Corollary}
\newtheorem{example}[define]{Example}

\newcommand{\proof}{{\it Proof. \quad}}
\newcommand{\qed}{\hbox{\rule[-2pt]{3pt}{6pt}}}

\begin{document}

\title{Dynamics and computation in functional shifts}

\author{Jun Namikawa\dag\footnote[3]{\textit{E-mail address}: jnamika@jaist.ac.jp} and Takashi Hashimoto\ddag}

\address{\dag\ddag\ School of Knowledge Science,
Japan Advanced Institute of Science and Technology,
Tatsunokuchi, Ishikawa, 923-1292, Japan}

\begin{abstract}
We introduce a new type of shift dynamics as an extended model of symbolic dynamics, and investigate the characteristics of shift spaces from the viewpoints of both dynamics and computation.
This shift dynamics is called a functional shift that is defined by a set of bi-infinite sequences of some functions on a set of symbols.
To analyze the complexity of functional shifts, we measure them in terms of topological entropy, and locate their languages in the Chomsky hierarchy.
Through this study, we argue that considering functional shifts from the viewpoints of both dynamics and computation give us opposite results about the complexity of systems.
We also describe a new class of shift spaces whose languages are not recursively enumerable.
\end{abstract}

\maketitle

\section{Introduction} \label{section:introduction}
We propose a new framework of shift dynamics, called functional shifts, which extends symbolic dynamics.
A functional shift is defined as a shift space that is a set of bi-infinite sequences of some functions on a set of symbols, while symbolic dynamics is usually defined as a set of bi-infinite sequences of finite symbols.
This functional shift generates another shift space, called a generated shift, as follows.
Consider a sequence of functions $(f_i)_{i \in \mathbb{Z}}$ contained in the functional shift.
The sequence of functions generates a sequence of symbols $(x_i)_{i \in \mathbb{Z}}$ determined by $x_{i+1} = f_i(x_i)$.
A set of such bi-infinite sequences of symbols is also a shift space.
Thus, this framework gives us a method to analyze the relationship among classes of shift spaces using the generative operation.

Introducing the framework of functional shifts allows us to consider the dynamic change of functions.
In traditional dynamical systems theory, the time dependence of a function governing change of states is not serious.
The reason for is that time in a dynamical system can be treated as an additional phase space variable, so that for any dynamical system we can describe it with a time independent function.
However, the time dependency of functions, i.e., the dynamic change of functions, has recently become a subject of interest in the investigation of dynamical systems, because there are many phenomena in which functions should be regarded as variable.
For example, in a population dynamics with species extinction and speciation, the degree of freedom dynamically changes \cite{Tokita1999,Tokita2003}.
In studying such phenomenon, since it is difficult to get an immutable evolution rule, we want to treat functions of the system as dynamic.
Therefore, considering dynamic changes of function is an important perspective for understanding complex systems.
We often call such dynamic changes meta-dynamics.

Several models have been proposed to study the dynamic change of functions \cite{Sato2000,Kataoka2000,Kataoka2001,Kataoka2003,Fontana1992,Fontana1994,Tsuda1994}.
Sato and Ikegami \cite{Sato2000} introduced switching map systems, in which maps to govern the evolution of the systems are dynamically switched with other maps in the system.
Kataoka and Kaneko \cite{Kataoka2000,Kataoka2001,Kataoka2003} investigated the evolution of a one-dimensional function $f_n$ defined by $f_{n+1} = (1 - \epsilon)f_n + \epsilon f_n \circ f_n$.
Studying dynamics in which functions vary in time using meta-dynamics can be important when considering systems evolution or learning.
Fontana \cite{Fontana1992} studied abstract chemical systems that are defined by a loop in which objects encode the functions that act on them.
For another instance, Tsuda \cite{Tsuda1994} proposed a switching map system as a model of the brain.
He has shown that a skew-product transformation can be considered as a framework describing meta-dynamics.
Functional shifts also can be represented by skew-product transformations.

The framework of functional shift has two major advantages.
One is the ability to directly compare dynamics with meta-dynamics, since both are represented by shift spaces.
The other is to be able to analyze both dynamical and computational characteristics, because this framework is an extension of symbolic dynamics.
Moreover, as a kind of meta-dynamics, we can discuss self-modifying systems in terms of functional shifts, in which the functions governing the dynamics of the system are used to change the functions themselves.

In this paper, we study functional shifts from the viewpoints of both dynamics and computation.
In recent years, several studies have focused on the relationship between dynamics and computation \cite{Moore1991,Siegelmann1995,Siegelmann1998,Siegelmann1999,Ben-Hur2002,Ben-Hur2003,Crutchfield1990,Lakdawala1996}.
The central idea in these studies is to regard the time evolution of dynamics as a computational process.
Based on a correspondence between the unpredictability of dynamical systems and the halting problem, Moore \cite{Moore1991} insists on the existence of dynamics that are more complex than chaos.
In addition, computational complexity for continuous time analog computation has been studied \cite{Siegelmann1998,Siegelmann1999,Ben-Hur2002,Ben-Hur2003}, in which the convergence of ordinary differential equations is interpreted as a process of computation.
The relationship between the complexity of dynamics and computation is, however, still unclear.
In this work we discuss the complexity of the dynamics of functional shifts in terms of topological entropy, which measures the diversity of orbits of dynamical systems, and also investigate the complexity of their computation in terms of the Chomsky hierarchy.
Through this study, we argue that dynamics and computation give us opposite results concerning the complexity of systems.

In an analysis of the complexity of computation in functional shifts, we prove that there exists a shift space whose language is not recursively enumerable (r.e.), even though the language of the functional shift that generates it is r.e.
Computational classes of sets to be beyond r.e.\ are related to analog computation in the interest of both dynamics and computation.
While Siegelmann \cite{Siegelmann1995} introduced analog shifts as a model of analog computation which is more powerful than the universal Turing machine, the dynamical features of such a powerful computation system are an open problem.
When we study dynamical systems modified by meta-dynamics, we discuss analog computation with functional shifts, and argue that the existence of dynamical behaviour that is more complex than the universal Turing machine should be taken into account.

This paper is organized as follows.
We first review some basic definitions of shift spaces, and define functional shifts in section \ref{section:definition}.
Next, in section \ref{section:entropy}, we investigate the property of entropy in functional shifts.
We show that the entropy of a functional shift gives the upper limit for that of a generated shift given by it, in order to study the relationship between dynamics and meta-dynamics (see theorem \ref{theorem:upper_limit_for_entropy_of_functional_shifts}).
In section \ref{section:computation}, we compare functional shifts with generated shifts, by focusing on how the language of a shift space belongs to the Chomsky hierarchy of formal languages.
We prove that any class of the languages of functional shifts is contained in that of the generated shifts given by them (see corollary \ref{corollary:class_of_FSs_subset_that_of_GSs} and theorem \ref{theorem:shift_of_finite_type_and_sofic}, \ref{theorem:cfg_to_cfg}, and \ref{theorem:language_of_generated_shift_beyond_r.e.}).
One of the most important results in this section is that there is a shift space whose language is not r.e., even though the language of a functional shift to generate the shift space is r.e.(see theorem \ref{theorem:language_of_generated_shift_beyond_r.e.}).
Finally, our results are discussed from the standpoints of dynamics, computation, meta-dynamics, and self-modifying systems.

\section{Definition} \label{section:definition}
Since we will study shift dynamics, we first give some definitions for shift spaces \cite{Lind1995}.
Let $\mathcal{A}$ be a nonempty finite set of symbols called an alphabet.
The full $\mathcal{A}$-shift (simply the full shift) is the collection of all bi-infinite sequences of symbols from $\mathcal{A}$.
Here such a sequence is denoted by $x = (x_i)_{i \in \mathbb{Z}}$ and the full $\mathcal{A}$-shift is denoted by
\begin{equation}
\mathcal{A}^{\mathbb{Z}} = \{x = (x_i)_{i \in \mathbb{Z}} |\forall i \in \mathbb{Z} ~ x_i \in \mathcal{A}\}.
\end{equation}
A block over $\mathcal{A}$ is a finite sequence of symbols from $\mathcal{A}$.
An $n$-block is simply a block of length $n$.
We write blocks without separating their symbols by commas or other punctuation, so that a typical block over $\mathcal{A} = \{a,b\}$ looks like $aababbabbb$.
A sequence of no symbols is called an empty block and denoted by $\epsilon$.
For any alphabet $\mathcal{A}$, we write $\mathcal{A}^*$ to denote the set of all blocks over $\mathcal{A}$.
If $x \in \mathcal{A}^{\mathbb{Z}}$ and $i \leq j$, then we denote a block of coordinates in $x$ from position $i$ to position $j$ by $x_{[i,j]} = x_i x_{i+1} \cdots x_{j}$.

Let $\sigma$ be a shift map on a full shift $\mathcal{A}^{\mathbb{Z}}$ defined by $\sigma((a_i)_{i \in \mathbb{Z}}) = (a_{i+1})_{i \in \mathbb{Z}}$ for any $a \in \mathcal{A}^{\mathbb{Z}}$.
A subset $X$ of $\mathcal{A}^{\mathbb{Z}}$ is called shift-invariant iff $\sigma(X) = X$.
Let $\mathtt{F}$, which we call the forbidden blocks, be a collection of blocks over $\mathcal{A}$.
For any such $\mathtt{F}$, define $X_{\mathtt{F}}$ to be the subset of sequences in $\mathcal{A}^{\mathbb{Z}}$ in which no block in $\mathtt{F}$ occurs.
A shift space is a subset $X$ of $\mathcal{A}^{\mathbb{Z}}$ such that $X = X_{\mathtt{F}}$ for some collection $\mathtt{F}$.
Note that if $X$ is a shift space then $X$ is shift-invariant, but a shift-invariant set is not necessarily a shift space (some examples are in Ref\cite{Lind1995}).
The set of all $n$-blocks that occur in points in $X$ is denoted by $\mathcal{B}_n(X)$, and the language of $X$ is the collection $\mathcal{B}(X) = \bigcup_{n=0}^{\infty}\mathcal{B}_{n}(X)$.
The language of a shift space determines the shift space.
Thus two shift spaces are equal iff they have the same language.

Suppose that $X$ is a shift space and $\mathcal{A}$ is an alphabet.
An ($m+n+1$)-block map $\Phi:\mathcal{B}_{m+n+1}(X) \rightarrow \mathcal{A}$ maps from allowed ($m+n+1$)-blocks in $X$ to symbols in $\mathcal{A}$.
A map $\phi:X \rightarrow \mathcal{A}^{\mathbb{Z}}$ defined by $y = \phi(x)$ with $y_i = \Phi(x_{i-m}x_{i-m+1} \cdots x_{i+n})$ is called a sliding block code induced by $\Phi$.
If $Y$ is a shift space over $\mathcal{A}$ and $\phi(X) \subset Y$, then we write $\phi:X \rightarrow Y$.
If a sliding block code $\phi:X \rightarrow Y$ is onto, $\phi$ is called a factor code.
A shift space $Y$ is a factor of $X$ if there is a factor code from $X$ onto $Y$.
A sliding block code $\phi:X \rightarrow Y$ is a conjugacy from $X$ to $Y$ if it is bijective.
Two shift spaces $X$ and $Y$ are conjugate (written $X \cong Y$) if there is a conjugacy between $X$ and $Y$.

Next, we define functional shifts and generated shifts, and explain the basic property of each.

\begin{define} \label{define:functional_shift}
{
\rm
Let $\mathcal{A}$ be a nonempty finite set, and $F$ be a set of maps on $\mathcal{A}$.
A \textit{functional shift} $\mathcal{F}$ is a shift space which is a subset of the full shift $F^{\mathbb{Z}}$.

A \textit{generated shift} $X_{\mathcal{F}}$ given by $\mathcal{F}$ is defined by
\begin{equation}
X_{\mathcal{F}} = \{(x_i)_{i \in \mathbb{Z}} \in \mathcal{A}^{\mathbb{Z}} |~ \exists (f_i)_{i \in \mathbb{Z}} \in \mathcal{F} ~ \forall i \in \mathbb{Z} ~ x_{i+1} = f_{i}(x_{i})\}.
\end{equation}
}
\end{define}

Although a generated shift is not required by our definition to be a shift space, it is always a shift space.
\begin{theorem} \label{theorem:generated_shifts_are_shift_spaces}
{
\rm
If $\mathcal{F}$ is a functional shift, $X_{\mathcal{F}}$ is a shift space.
}
\end{theorem}
\proof
Let
\begin{equation}
Y_n = \{x \in \mathcal{A}^n | ~ \forall f \in \mathcal{B}_n(\mathcal{F}) ~ \exists i ~ x_{i+1} \neq f_i(x_i)\}
\end{equation}
and $\mathtt{F} = \bigcup_{n \in \mathbb{N}} Y_n$.
Suppose that $X$ is a shift space which can be described by the collection $\mathtt{F}$ of forbidden blocks.
If $x \in X_{\mathcal{F}}$, then $x \in X$ because every block in $\mathtt{F}$ does not occur in $x$.
Thus $X_{\mathcal{F}} \subset X$.

Conversely if $x \in X$, then $x_{[-n,n]} \not\in \mathtt{F}$ for all $n$ because $\mathtt{F}$ is a set of blocks never occurring in points in $X$.
Therefore,
\begin{equation}
\forall n \in \mathbb{N} ~ \exists f \in \mathcal{B}_{2n}(\mathcal{F}) ~ \forall i ~ x_{[-n,n]i+1} = f_i(x_{[-n,n]i}),
\end{equation}
then
\begin{equation}
\exists f \in \mathcal{F} ~ \forall i \in \mathbb{Z} ~ x_{i+1} = f_{i}(x_{i}),
\end{equation}
so that $x \in X_{\mathcal{F}}$.
Accordingly $X \subset X_{\mathcal{F}}$.
Hence $X = X_{\mathcal{F}}$ and $X_{\mathcal{F}}$ is a shift space.
\hfill \qed
\newline

By theorem \ref{theorem:generated_shifts_are_shift_spaces}, a functional shift is regarded as a rule to generate a shift space.
Since any functional shift is also a shift space, we can compare functional shifts with generated shifts by using the properties of shift spaces.

The following examples are instances of functional shifts which generate shift spaces.
\begin{example}[Full shifts] \label{example:full_shift}
{
\rm
Let $\mathcal{A} = \{0,1\}$ and $F = \{f,g\}$ be a set of maps such that 
\begin{center}
\begin{tabular}{|ccc|}
\hline
$x$ & $f(x)$ & $g(x)$ \\
\hline
$0$ & $1$ & $0$ \\
$1$ & $0$ & $1$ \\
\hline
\end{tabular}
\end{center}
If a functional shift $\mathcal{F}$ is equal to $F^{\mathbb{Z}}$, then $X_{\mathcal{F}}$ is the full shift $\mathcal{A}^{\mathbb{Z}}$.
}
\end{example}

\begin{example}[Golden mean shift] \label{example:golden_mean_shift}
{
\rm
Let $X$ be the set of all binary sequences with no two $1$'s next to each other, so that $X = X_{\mathtt{F}}$, where $\mathtt{F} = \{11\}$.
This is called the golden mean shift.

Let $\mathcal{A} = \{0,1\}$ and $F = \{f,g\}$ be a set of maps such that 
\begin{center}
\begin{tabular}{|ccc|}
\hline
$x$ & $f(x)$ & $g(x)$ \\
\hline
$0$ & $1$ & $0$ \\
$1$ & $0$ & $0$ \\
\hline
\end{tabular}
\end{center}
If a functional shift $\mathcal{F}$ is equal to $F^{\mathbb{Z}}$, then $X_{\mathcal{F}}$ is the set of all binary sequences not to contain contiguous $1$'s.
Thus $X_{\mathcal{F}}$ is equal to the golden mean shift $X$.
}
\end{example}

\begin{example}[Sturmian shifts] \label{example:sturmian_shift}
{
\rm
Consider the circle map
\begin{equation} \label{equation:circle_map}
T(x) = x + \alpha \bmod 1
\end{equation}
with irrational $\alpha \in [0,1]$.
Let $S \subset \{0,1\}^{\mathbb{Z}}$ be a set
\begin{equation} \label{equation:sturmian_shift}
S = \{s \in \{0,1\}^{\mathbb{Z}} | x \in [0,1), \forall n \in \mathbb{Z} ~ s_n = \lfloor T^n(x)/\alpha \rfloor \},
\end{equation}
where $\lfloor x \rfloor$ is the integer part of $x$.
$S$ is not necessarily closed, but it is shift-invariant, and so its closure $X_{\alpha} = \mbox{Cl}(S)$ is a shift space, called a Sturmian shift.

Let $F = \{f,g\}$ be a set of maps given by
\begin{center}
\begin{tabular}{|ccc|}
\hline
$x$ & $f(x)$ & $g(x)$ \\
\hline
$0$ & $1$ & $1-\lfloor 2\alpha \rfloor$ \\
$1$ & $0$ & $1-\lfloor 2\alpha \rfloor$ \\
\hline
\end{tabular}
\end{center}
and $X_{\alpha}$ is a Sturmian shift with irrational $\alpha$.
Here $\mathcal{F} = \phi(X_{\alpha})$, where a 2-block map $\Phi:\{0,1\}^2 \rightarrow F$ is defined by 
\begin{equation}
\Phi(x) = \left\{ \begin{array}{ll}
   f  & \mbox{if $x = 01$ or $x = 10$}, \\
   g  & \mbox{otherwise}, \\
\end{array} \right.
\end{equation}
and a map $\phi:X_{\alpha} \rightarrow F^{\mathbb{Z}}$ is a sliding block code induced by $\Phi$.
Since $\alpha$ is irrational, $00$ or $11$ must appear in $x$, so that for any $h \in \mathcal{F}$ there is a unique sequence $x \in X_{\alpha}$ such that $\phi(x) = h$ and $x_{n+1} = h_n(x_n)$ for all $n \in \mathbb{Z}$.
Therefore this functional shift $\mathcal{F}$ satisfies $\mathcal{F} \cong X_{\alpha}$ and $X_{\mathcal{F}} = X_{\alpha}$.
}
\end{example}

\section{Entropy} \label{section:entropy}
This section will describe the properties of the relationship between functional shifts and generated shifts by analyzing the entropy for those shifts.

The entropy of a shift space $X$ is defined by
\begin{equation} \label{equation:entropy_of_shift_space}
h(X) = \lim_{n \rightarrow \infty}\frac{1}{n}\log_2|\mathcal{B}_{n}(X)|.
\end{equation}
Note that $\log_2 |\mathcal{A}|$ is the upper limit of $h(X)$ for any $X \subset \mathcal{A}^{\mathbb{Z}}$.
The entropy of $X$ is a measure of the growth rate of the number of $n$-blocks occurring in points in $X$.
Furthermore, if a distance function $d$ of $X$ is determined by 
\begin{equation}
d(x,y) = \left\{ \begin{array}{ll}
	2^{-|k|}	& \qquad \mbox{if $x_{k} \neq y_{k}$ and $x_{i} = y_{i}$ for $-|k| < i < |k|$},  \\
	0	& \qquad \mbox{if $x = y$}, \\
	\end{array} \right.
\end{equation}
then the entropy of $X$ is equal to the topological entropy of the shift map on the metric space $(X,d)$ \cite{Lind1995}.
Hence we regard the entropy of a shift space as the topological entropy.

It is known that the topological entropy is an indicator of the complexity of the dynamics and that it is invariant under topological conjugacy.
The existence of positive topological entropy implies that a system is chaotic, because the topological entropy measures the mixing rate of the global orbit structure of the system.

The following theorem is an important result of the entropy of functional shifts.
\begin{theorem} \label{theorem:upper_limit_for_entropy_of_functional_shifts}
{
\rm
If $\mathcal{F}$ is a functional shift, then $h(X_{\mathcal{F}}) \leq h(\mathcal{F})$.
}
\end{theorem}
\proof
Let $\varphi_n:\mathcal{B}_n(\mathcal{F}) \rightarrow 2^{\mathcal{A}^n}$ be defined by
\begin{equation}
\varphi_n(f) = \{x \in \mathcal{B}_{n}(X_{\mathcal{F}}) | ~ \exists a \in \mathcal{A} ~ x_0 = f_0(a) \land \forall i ~ x_{i+1} = f_{i+1}(x_{i}) \},
\end{equation}
where $2^X$ denotes the power set of $X$.
It is clear that $|\varphi_n(f)| \leq |\mathcal{A}|$ for all $f \in \mathcal{B}_n(\mathcal{F})$ and 
\begin{equation}
\mathcal{B}_{n}(X_{\mathcal{F}}) \subset \bigcup_{f \in \mathcal{B}_n(\mathcal{F})}\varphi_n(f).
\end{equation}
Thus $|\mathcal{B}_{n}(X_{\mathcal{F}})| \leq |\mathcal{B}_n(\mathcal{F})||\mathcal{A}|$.
Accordingly,
\begin{eqnarray}
h(X_{\mathcal{F}}) & = & \lim_{n \rightarrow \infty}\frac{1}{n}\log_2|\mathcal{B}_{n}(X_{\mathcal{F}})| \nonumber \\
 & \leq & \lim_{n \rightarrow \infty}\frac{1}{n}\log_2(|\mathcal{B}_{n}(\mathcal{F})||\mathcal{A}|) \nonumber \\
 & = & \lim_{n \rightarrow \infty}\frac{1}{n}\log_2|\mathcal{B}_{n}(\mathcal{F})| \nonumber \\
  & = & h(\mathcal{F}).
\end{eqnarray}
Hence $h(X_{\mathcal{F}}) \leq h(\mathcal{F})$.
\hfill \qed

\begin{example}
{
\rm
Let $\mathcal{F}$ be a functional shift given in example \ref{example:golden_mean_shift}.
Here $h(\mathcal{F}) = \log_2 2$ and $h(X_{\mathcal{F}}) = \log_2 \frac{1+\sqrt{5}}{2}$ (this derivation formula is in Ref \cite{Lind1995}).
Thus $h(X_{\mathcal{F}}) < h(\mathcal{F})$.
}
\end{example}

The key point in the proof of theorem \ref{theorem:upper_limit_for_entropy_of_functional_shifts} is to satisfy a condition $|\mathcal{B}_{n}(X_{\mathcal{F}})| \leq |\mathcal{B}_n(\mathcal{F})||\mathcal{A}|$ for any $n \in \mathbb{N}$.
The condition implies that for any sequence of symbols $x \in X_{\mathcal{F}}$ there exists one or more sequences of functions $f \in \mathcal{F}$ which are restricted by $x_{n+1} = f_n(x_n)$ for all $n \in \mathbb{Z}$.
Since the plural sequences (in some case, it is infinitely) in $\mathcal{F}$ can correspond to a sequence in $X_{\mathcal{F}}$, the entropy of $\mathcal{F}$ is the same as or larger than that of $X_{\mathcal{F}}$.

If $F$ is a set of maps on $\mathcal{A}$ such that 
\begin{equation}
\forall f,g \in F ~ f \neq g \Rightarrow \forall x \in \mathcal{A} ~ f(x) \neq g(x),
\end{equation}
then every functional shift $\mathcal{F} \subset F^{\mathbb{Z}}$ satisfies $h(X_{\mathcal{F}}) = h(\mathcal{F})$, because $|\mathcal{B}_{n}(X_{\mathcal{F}})| \geq |\mathcal{B}_n(\mathcal{F})|$ for all $n \in \mathbb{N}$.
Hence
\begin{equation} \label{equation:redundancy_functions_set}
\exists f,g \in F ~ \exists x \in \mathcal{A} ~ f \neq g \land f(x) = g(x)
\end{equation}
is a necessary condition in order to realize $h(X_{\mathcal{F}}) < h(\mathcal{F})$.
For example, if $|\mathcal{A}| < |F|$ and $\mathcal{F} = F^{\mathbb{Z}}$, then $F$ satisfies equation (\ref{equation:redundancy_functions_set}) and $h(X_{\mathcal{F}}) < h(\mathcal{F})$.
However, there exists a case in which $F$ satisfies equation (\ref{equation:redundancy_functions_set}) and $h(X_{\mathcal{F}}) = h(\mathcal{F})$.
The example \ref{example:sturmian_shift} is one of such instances, because $\mathcal{F}$ and $X_{\mathcal{F}}$ are conjugate and the entropy is invariant under conjugacy.

From theorem \ref{theorem:upper_limit_for_entropy_of_functional_shifts}, we may consider that the degree of complexity of a functional shift $\mathcal{F}$ is greater than that of $X_{\mathcal{F}}$ from the viewpoint of dynamics.
However, satisfying such a relationship is not necessarily required in the computational point of view.
The next section turns to the computational power of shift spaces, and compares the languages of functional shifts with those of generated shifts.

\section{Computation in functional shifts} \label{section:computation}
In section \ref{section:entropy}, we compared the complexity of functional shifts with generated shifts based on entropy.
Entropy measures the exponential growth rate of the number of orbits distinguished in limited accuracy, i.e., it represents sensitivity to the initial conditions of a dynamical system.
On the other hand, there is the complexity of languages given by the Chomsky hierarchy which differs from that of entropy.
The complexity described by the Chomsky hierarchy is based on the memory size of the automata that recognize languages (see \ref{section:appendix_a}).
In this sense languages are classified into four classes: regular languages which do not need any memory; context free languages which have just a stack; context sensitive languages which have a storage capacity proportional to the input word length; and type-0 languages which have unlimited memory.

The memory size of an automaton is deeply related to the long-range correlation and unpredictability of a dynamical system.
Badii and Politi have discussed the relationship between memory size and the properties of dynamical systems with some examples of physical systems corresponding to formal languages in the Chomsky hierarchy \cite{Badii1997}.
For example, random walks with two reflecting barriers are dynamics whose languages are regular, because the domain surrounded by two barriers can be considered to express a finite state.
Those dynamics can be described by the Markov graph, so they are stationary and ergodic.
Random walks with one reflecting barrier are dynamics whose languages are context free, because the domain on the semi-lattice that makes one barrier the starting point can be considered to express a stack.
The fact that there is no restriction about distance from a barrier brings long-range correlation to those random walks.
For another example, self-avoiding random walks are dynamics whose languages are more complex than context free languages.
Some dynamics whose languages are not context free have strong unpredictability.
In section \ref{section:discussion}, we discuss this property in detail.

In this section, we compare functional shifts with generated shifts, by focusing on how the language of a shift space belongs to the Chomsky hierarchy of formal languages.
Notice that not every collection of blocks is the language of a shift space.
Namely, if $X$ is a shift space and $w \in \mathcal{B}(X)$, then
\begin{ul}{0.2}
\item every subblock of $w$ belongs to $\mathcal{B}(X)$, and
\item there are nonempty blocks $u$ and $v$ in $\mathcal{B}(X)$ such that $uwv \in \mathcal{B}(X)$.
\end{ul}

Given a class of languages of functional shifts $\mathscr{F}$, a class of languages of generated shifts $\mathscr{G}$ is given by $\mathscr{G} = \{ \mathcal{B}(X_{\mathcal{F}}) | ~ \mathcal{B}(\mathcal{F}) \in \mathscr{F}, \mbox{$\mathcal{F}$ is a functional shift}\}$.
Now we consider the inclusion relation between $\mathscr{F}$ and $\mathscr{G}$.
From the properties of entropy in functional shifts, we may consider that $\mathscr{F}$ is at least as complex as $\mathscr{G}$ from the dynamical viewpoint, because for any shift space $X$ whose language is in $\mathcal{G}$ there exists a functional shift $\mathcal{F}$ whose language is in $\mathscr{F}$ satisfying $X_{\mathcal{F}} = X$, so that $h(X) \leq h(\mathcal{F})$.
Thus, if the complexity of entropy corresponds to that of computation, we shall expect $\mathscr{G}$ as the subclass of $\mathscr{F}$.
It is, however, known that the complexities of entropy and computation generally do not correspond.
For instance, consider the language of a periodic shift space which only contains periodic sequences and that of a full shift.
Both are contained in a class of regular languages which is the lowest computation class in the Chomsky hierarchy.
However, the former has minimal entropy, and the latter has maximal entropy.
The fact that shift spaces with different entropy belong to the same computational class makes it generally difficult to clarify the relationship between entropy and the complexity of computation.
Hence it is worthwhile to investigate this relationship using functional shifts.
We will show the inclusion relation between $\mathscr{F}$ and $\mathscr{G}$ in the case where $\mathscr{F}$ and $\mathscr{G}$ belong to the Chomsky hierarchy, and discuss the complexity of dynamics and computation with these results.

We first prove the following theorem to be a basic principle of functional shifts.
\begin{theorem} \label{theorem:functional_shift_is_equal_to_generated_shift}
{
\rm
For any shift space $X$ over $\mathcal{A}$, there is a functional shift $\mathcal{F}$ over $F$ and a $1$-block map $\Phi:\mathcal{A} \rightarrow F$ such that
\begin{ul}{0.2}
\item $X_{\mathcal{F}} = X$,
\item $\Phi$ is a one-to-one mapping,
\item $\phi$ induced by $\Phi$ satisfies $\phi(X) = \mathcal{F}$.
\end{ul}
}
\end{theorem}
\proof
For any $a \in \mathcal{A}$, we define a function $f_a: \mathcal{A} \rightarrow \mathcal{A}$ by
\begin{equation}
\forall x \in \mathcal{A} ~ f_a(x) = a.
\end{equation}
Let $F = \{f_a | a \in \mathcal{A}\}$, and $\Phi$ be defined by
\begin{equation}
\forall a \in \mathcal{A} ~ \Phi(a) = f_a.
\end{equation}
Clearly $\Phi$ is a one-to-one mapping, so that $\phi(X)$ is a shift space.
Now a functional shift $\mathcal{F}$ is defined by $\mathcal{F} = \phi(X)$.
Then
\begin{eqnarray}
x \in X_{\mathcal{F}}  & \Leftrightarrow & \exists f \in \mathcal{F} ~ \forall i \in \mathbb{Z} ~ x_{i+1} = f_i(x_i) \nonumber \\
 & \Leftrightarrow & \exists f \in \mathcal{F} ~ \forall i \in \mathbb{Z} ~ \Phi(x_{i+1}) = f_i \nonumber \\
 & \Leftrightarrow & x \in X.
\end{eqnarray}
Thus $X_\mathcal{F} = X$.
\hfill \qed
\newline

From theorem \ref{theorem:functional_shift_is_equal_to_generated_shift} we can get the next corollary.
\begin{corollary} \label{corollary:class_of_FSs_subset_that_of_GSs}
{
\rm
Let $\mathscr{F}$ be a class of languages of shift spaces.
Suppose that $\mathscr{G}$ is a class of languages of generated shifts given by the functional shifts whose languages belong to $\mathscr{F}$.
Then $\mathscr{F} \subset \mathscr{G}$.
}
\end{corollary}
\proof
Let $L$ be a language in $\mathscr{F}$, and $X$ be a shift space defined by $\mathcal{B}(X) = L$.
By theorem \ref{theorem:functional_shift_is_equal_to_generated_shift}, there is a functional shift $\mathcal{F}$ such that $\mathcal{B}(X) = \mathcal{B}(\mathcal{F})$ and $X_{\mathcal{F}} = X$.
Then $L \in \mathscr{G}$.
\hfill \qed
\newline

From corollary \ref{corollary:class_of_FSs_subset_that_of_GSs}, any class of the languages of functional shifts is contained in that of generated shifts given by them.
However, there is still the open problem of whether there exists a class of languages of functional shifts which is proper subset of languages of shift spaces generated by the functional shifts.
We can study the relationship between the languages of functional shifts and those of generated shifts to bring this problem into focus.

Hereafter, the next lemma is the key ingredient in each proof.
\begin{lemma} \label{lemma:condition_of_word_adscription}
{
\rm
Let
\begin{equation}
\mathcal{D}_{\mathcal{F}}(n, x) = \{y = ax | a \in \mathcal{A}^{n}, \exists f \in \mathcal{B}_{|y|-1}(\mathcal{F}) ~ \forall i ~ y_{i+1} = f_i(y_i) \}.
\end{equation}
Then $x \in \mathcal{B}(X_{\mathcal{F}})$ iff $\lim_{n \rightarrow \infty}\mathcal{D}_{\mathcal{F}}(n, x) \neq \emptyset$.
}
\end{lemma}
\proof
Suppose that $x \in \mathcal{B}(X_{\mathcal{F}})$, and $y \in X_{\mathcal{F}}$ having subblock $x$.
There is a bi-infinite sequence $f \in \mathcal{F}$ such that $y_{i+1} = f_i(y_i)$ for any $i \in \mathbb{Z}$.
Since $x$ is a subblock of $y$, an integer $k$ such as $x_1 = y_k$ exists.
Hence $y_{k-n} \cdots y_{k-1}x_1\cdots x_{|x|} \in \mathcal{D}_{\mathcal{F}}(n, x)$ for any $n \in \mathbb{N}$, so that $\lim_{n \rightarrow \infty}\mathcal{D}_{\mathcal{F}}(n, x) \neq \emptyset$.

Conversely, suppose that $\lim_{n \rightarrow \infty}\mathcal{D}_{\mathcal{F}}(n, x) \neq \emptyset$.
Then there is an infinite sequence $y = \cdots a_{-1}a_{0}x_1 \cdots x_{|x|}$ (by Koenig's lemma) and a bi-infinite sequence $f \in \mathcal{F}$ such that $y_{i+1} = f_i(y_i)$ for $-\infty < i < |x|$.
If a bi-infinite sequence $z$ is defined by
\begin{equation}
z_i = \left\{ \begin{array}{ll}
	y_i	& \qquad \mbox{if $i \leq |x|$}, \\
	f_{i-1}(z_{i-1})	& \qquad \mbox{otherwise}, \\
	\end{array} \right.
\end{equation}
then $z \in X_{\mathcal{F}}$ because $z_{i+1} = f_i(z_i)$ for any $i \in \mathbb{Z}$.
Since $x$ is a subblock of $z$, $x \in \mathcal{B}(X_{\mathcal{F}})$.
\hfill \qed

\subsection{Shifts of finite type and sofic shifts}
Here we study the case in which a functional shift is a shift of finite type or a sofic shift.
We first define shifts of finite type and sofic shifts.

A shift of finite type is a shift space that can be described by a finite set of forbidden blocks.
Although shifts of finite type are the simplest shifts, they are significant in the dynamical systems theory.
If a dynamical system is hyperbolic, then the system has a Markov partition and a topological conjugacy to a shift of finite type.

Sofic shifts are defined by using graphs, called labeled graphs, whose edges are assigned labels.
A graph $G$ consists of a finite set $\mathcal{V} = \mathcal{V}(G)$ of vertices (or states) together with a finite set $\mathcal{E} = \mathcal{E}(G)$ of edges.
Each edge $e \in \mathcal{E}$ starts at a vertex denoted by $i(e) \in \mathcal{V}(G)$ and terminates at a vertex $t(e) \in \mathcal{V}(G)$ (which can be the same as $i(e)$).
Equivalently, the edge $e$ has an initial state $i(e)$ and a terminal state $t(e)$.
A labeled graph $\mathcal{G}$ is a pair $(G,\mathcal{L})$, where $G$ is a graph with edge set $\mathcal{E}$, and the labeling $\mathcal{L}:\mathcal{E} \rightarrow \mathcal{A}$ assigns each edge $e$ of $G$ to a label $\mathcal{L}(e)$ in $\mathcal{A}$.
Let $X_{\mathcal{G}}$ be denoted by
\begin{equation}
X_{\mathcal{G}} = \{x \in \mathcal{A}^{\mathbb{Z}} |~ \exists e \in \mathcal{E}^{\mathbb{Z}} ~ \forall i \in \mathbb{Z} ~ t(e_i) = i(e_{i+1}) \land x_i = \mathcal{L}(e_i)\}.
\end{equation}
A subset $X$ of the full shift $\mathcal{A}^{\mathbb{Z}}$ is a sofic shift if $X = X_{\mathcal{G}}$ for some labeled graph $\mathcal{G}$.
Since a labeled graph is regarded as a state diagram of a finite state automaton, the language of a sofic shift is regular.

It is known that a shift space is sofic iff it is a factor of a shift of finite type \cite{Lind1995}.
Since an identity function on a shift space is a factor code, shifts of finite type are sofic.
Moreover, the class of sofic shifts is larger than that of shifts of finite type, because not all sofic shifts have finite type.
For example, the even shift, which can be described by the collection $\{10^{2n+1}1|n \geq 0\}$ of forbidden blocks, is a sofic shift that does not have finite type.

Let us prove the following theorems as the case in which functional shifts are shifts of finite type or sofic shifts.
\begin{theorem} \label{theorem:shift_of_finite_type_and_sofic}
{
\rm
\begin{ol}{0.2}
\item If $X$ is a sofic shift, then there is a functional shift $\mathcal{F}$ such that $\mathcal{F}$ has finite type and $X = X_{\mathcal{F}}$.
\item If a functional shift $\mathcal{F}$ is sofic, then $X_{\mathcal{F}}$ is sofic (so that if a functional shift $\mathcal{F}$ has finite type, $X_{\mathcal{F}}$ is sofic).
\end{ol}
}
\end{theorem}
\proof
(1) Suppose that $X$ is a sofic shift over $\mathcal{A}$, and $\mathcal{G} = (G,\mathcal{L})$ is a labeled graph such that $X = X_{\mathcal{G}}$.
If $a$ is an edge of $G$, then $f_a:\mathcal{A} \cup \mathcal{E} \cup \{\delta\} \rightarrow \mathcal{A} \cup \mathcal{E} \cup \{\delta\}$ (where $\mathcal{A} \cap \mathcal{E} = \emptyset$ and $\delta \not\in \mathcal{A} \cup \mathcal{E}$) is defined by
\begin{equation}
f_a(x) = \left\{ \begin{array}{ll}
	\delta	& \qquad \mbox{if $x = a$}, \\
	\mathcal{L}(a)	& \qquad \mbox{otherwise}. \\
	\end{array} \right.
\end{equation}
Let $F = \{f_a | ~ a \in \mathcal{E}\}$ and $\mathtt{F} = \{f_af_b \in F^2 | ~ t(a) \neq i(b)\}$.
Recall that $i(a)$ is an initial state and $t(a)$ is a terminal state of $a$.
If a functional shift $\mathcal{F}$ over $F$ can be described by $\mathtt{F}$, then $X_{\mathcal{F}} = X$.
Since $\mathtt{F}$ is a finite set, $\mathcal{F}$ is a shift of finite type.

(2) Suppose that $\mathcal{F}$ is a sofic shift over $F$ which is a set of maps on $\mathcal{A}$, and $F'$ is a collection of maps on $F$.
By (1), there is a functional shift $\mathcal{F}'$ over $F'$ such that $\mathcal{F}'$ has finite type and $X_{\mathcal{F}'} = \mathcal{F}$.
Let 
\begin{equation} \label{equation:product_of_functional_shift_and_generated_shift}
\hspace*{-4em} X = \{\langle x,f,g \rangle \in (\mathcal{A} \! \times F \! \times F')^{\mathbb{Z}} | ~ g \in \mathcal{F}', \forall i \in \mathbb{Z} ~ x_{i+1} = f_i(x_i) \land f_{i+1} = g_i(f_i)\},
\end{equation}
be a set of elements which are sequences of 3-tuples $(\cdots \langle x_0,f_0,g_0 \rangle \langle x_1,f_1,g_1 \rangle \cdots)$, and $\mathtt{F}$ be a finite set of forbidden blocks such that $X_{\mathtt{F}} = \mathcal{F}'$.
Then
\begin{eqnarray}
\tilde{\mathtt{F}} & = & \{\langle x,f,g \rangle \in (\mathcal{A} \times F \times F')^{2} | ~ x_1 \neq f_0(x_0) \lor f_1 \neq g_0(f_0)\} \nonumber \\
& & \cup ~ \{\langle x,f,g \rangle \in (\mathcal{A} \times F \times F')^{*} | ~ g \in \mathtt{F}\}
\end{eqnarray}
is a set of forbidden blocks such that $X_{\tilde{\mathtt{F}}} = X$.
Here, $\tilde{\mathtt{F}}$ is a finite set because $\mathtt{F}$ is finite.
Suppose that $\Phi:(\mathcal{A} \times F \times F') \rightarrow \mathcal{A}$ is a $1$-block map such that $\Phi(\langle x,f,g \rangle) = x$.
Since a sliding block code $\phi:X \rightarrow X_{\mathcal{F}}$ induced by $\Phi$ is onto, $\phi$ is a factor code.
If a shift space is a factor of a shift of finite type, then it is sofic.
Thus $X_{\mathcal{F}}$ is a sofic shift.
\hfill \qed

The next is an instance satisfying theorem \ref{theorem:shift_of_finite_type_and_sofic} (1).

\begin{example} \label{example:even_shift_generated_by_functional_shift}
{
\rm
Let $\mathcal{A} = \{0,1\}$, and $F = \{f_a, f_b, f_c\}$ be a set of functions on $\mathcal{A}$ such that
\begin{center}
\begin{tabular}{|cccc|}
\hline
$x$ & $f_a(x)$ & $f_b(x)$ & $f_c(x)$ \\
\hline
$0$ & $1$ & $0$ & $0$ \\
$1$ & $1$ & $0$ & $1$ \\
\hline
\end{tabular}
\end{center}
If a functional shift $\mathcal{F}$ can be described by a set of forbidden blocks $\mathtt{F} = \{f_af_c, f_bf_a, f_bf_b, f_cf_c\}$, then $\mathcal{F}$ is a shift of finite type and $X_{\mathcal{F}}$ is the even shift.
}
\end{example}

\subsection{Context free languages}
This subsection studies the case in which the language of a functional shift is a context free language.
The shift dynamics on some shift spaces with languages that are context free have long-range correlations, because stacks can hold memories infinitely.
Moreover, if the shift spaces are probability measure spaces, the phase transition often appears in this dynamics \cite{Badii1997}.

We begin by proving that if the language of a functional shift $\mathcal{F}$ is context free, then there is a number $p \in \mathbb{N}$ such that for any $x \in \mathcal{A}^*$ $\mathcal{D}_{\mathcal{F}}(p, x) \neq \emptyset$ iff $\lim_{n \rightarrow \infty}\mathcal{D}_{\mathcal{F}}(n, x) \neq \emptyset$, by using the pumping lemma\footnote[1]{In the theory of formal languages, the pumping lemma provides necessary conditions for languages to be context free. The pumping lemma for context free languages is as follows: if language $L$ is context free, then there is a natural number $p$ such that if $r = uvwxy \in L$ and $|r| > p$ then $|vx| \geq 1$, $|vwx| \leq p$, and for any $i \geq 0$, $uv^iwx^iy \in L$.}.
Next we prove that if the language of $\mathcal{F}$ is context free, then that of $X_{\mathcal{F}}$ is so.

\begin{lemma} \label{lemma:pumping_lemma}
{
\rm
Suppose that $\mathcal{F}$ is a functional shift and $\mathcal{B}(\mathcal{F})$ is a context free language.
There is a natural number $p$ such that
\begin{equation}
\forall x \in \mathcal{A}^* \quad \mathcal{D}_{\mathcal{F}}(p, x) \neq \emptyset \Leftrightarrow \lim_{n \rightarrow \infty}\mathcal{D}_{\mathcal{F}}(n, x) \neq \emptyset.
\end{equation}
}
\end{lemma}
\proof
Let $G = \{N,F,P,S\}$ be a context free grammar such that $\mathcal{B}(\mathcal{F}) = L(G)$, where $L(G)$ denotes a formal language generated by $G$.
Suppose, without loss of generality, that $G$ is in Chomsky normal form\footnote{A formal grammar $G = (V_N,V_T,P,S)$ is in Chomsky normal form iff all productions are of the form $A \rightarrow BC$ or $A \rightarrow a$, where $A,B,C \in V_N$ and $a \in V_T$. Every formal grammar in Chomsky normal is context free, and conversely, every context free grammar that does not generate an empty string can be transformed into an equivalent one which is in Chomsky normal form.}.
Now a formal grammar $G' = \{N', \mathcal{A}, P',S'\}$ is defined as follows.
A set $N'$ of nonterminal symbols is equal to $\{A_{ab} | A \in N, a, b \in \mathcal{A}\}$.
$P'$ is a set of productions determined by the following rules:
\begin{ul}{0.2}
\item $S' \rightarrow aS_{ab} \in P'$ \quad for any $a, b \in \mathcal{A}$;
\item $A_{ab} \rightarrow B_{ac}C_{cb} \in P'$ iff $A \rightarrow BC \in P$, where $A,B,C \in N$ and $a,b,c \in \mathcal{A}$;
\item $A_{ab} \rightarrow b \in P'$ iff $A \rightarrow f \in P$ and $f(a) = b$, where $A \in N$, $f \in F$, and $a ,b \in \mathcal{A}$.
\end{ul}
Clearly, $G'$ is a context free grammar, furthermore, 
\begin{equation}
L(G') = \bigcup_{x \in \mathcal{A}^*}\mathcal{D}_{\mathcal{F}}(0,x) = \bigcup_{x \in \mathcal{A}^*}\bigcup_{n \in \mathbb{N}}\mathcal{D}_{\mathcal{F}}(n,x)
\end{equation}
because $x \in L(G')$ iff there is a block $f \in L(G) = \mathcal{B}(\mathcal{F})$ such that $|f| = |x|-1$ and $x_{i+1} = f_i(x_i)$ for $1 \leq i < |x|$.

From the pumping lemma, there is a natural number $p$ such that if $r = uvwxy \in L(G')$ and $|r| > p$ then 
\begin{ul}{0.2}
\item $|vx| \geq 1$,
\item $|vwx| \leq p$,
\item for any $i \geq 0$, $uv^iwx^iy \in L(G')$.
\end{ul}
Hence if $rs \in L(G')$, i.e., $\mathcal{D}_{\mathcal{F}}(p, s) \neq \emptyset$, then $\mathcal{D}_{\mathcal{F}}(p+|vx|n,s) \neq \emptyset$ for all $n \in \mathbb{N}$.
Note that if $\mathcal{D}_{\mathcal{F}}(n, s) \neq \emptyset$ and $m \leq n$ then $\mathcal{D}_{\mathcal{F}}(m, s) \neq \emptyset$.
Therefore, $\mathcal{D}_{\mathcal{F}}(p, s) \neq \emptyset$ iff $\lim_{n \rightarrow \infty}\mathcal{D}_{\mathcal{F}}(n, s) \neq \emptyset$.
\hfill \qed
\newline

Since we use a nondeterministic pushdown automaton (NPDA) to prove the next theorem, we give the formal definition of NPDA.
A NPDA is a 6-tuple $M = \{Q, \Sigma, \Gamma, \delta, q_0, Z, E\}$, where $Q$ is a finite set of states, $\Sigma$ is an alphabet (defining what set of input strings the automaton operates on), $\Gamma$ is a stack alphabet (specifying the set of symbols that can be pushed onto the stack), $\delta:Q \times (\Sigma \cup \{\epsilon\}) \times \Gamma \rightarrow 2^{Q \times \Gamma^*}$ is a transition function, $q_0 \in Q$ is a starting state, $Z \in \Gamma$ is a starting stack symbol, and $E \subset Q$ is a set of final (or accepting) states.

Given a NPDA $M = \{Q, \Sigma, \Gamma, \delta, q_0, Z, E\}$, the relation $\underset{M}{\vdash} \subset Q \times \Sigma^* \times \Gamma^* \times Q \times \Sigma^* \times \Gamma^*$ is defined by
\begin{equation}
(p, \beta) \in \delta(q,a,A) \Leftrightarrow (q, aw, A\gamma) \underset{M}{\vdash} (p, w, \beta\gamma),
\end{equation}
and the reflexive and transitive closure $\underset{M}{\vdash}$ is denoted by $\overset{*}{\underset{M}{\vdash}}$.
$M$ accepts an input string $x$ if there are $p \in E$ and $\gamma \in \Gamma^*$ such that $(q_0, x, Z) \overset{*}{\underset{M}{\vdash}} (p, \epsilon, \gamma)$.
The language recognized by $M$ is the set
\begin{equation}
L(M) = \{x \in \Sigma^* | ~ \exists p \in E ~ \exists \gamma \in \Gamma^* ~ (q_0, x, Z) \overset{*}{\underset{M}{\vdash}} (p, \epsilon, \gamma) \}.
\end{equation}
It is known that a language $L$ is a context free language iff $L = L(M)$ for some NPDA $M$.

\begin{theorem} \label{theorem:cfg_to_cfg}
{
\rm
If $\mathcal{F}$ is a functional shift and $\mathcal{B}(\mathcal{F})$ is a context free language, then $\mathcal{B}(X_{\mathcal{F}})$ is also a context free language.
}
\end{theorem}
\proof
We will construct a NPDA which can recognize the language $\mathcal{B}(X_{\mathcal{F}})$.

Since $\mathcal{B}(\mathcal{F})$ is a context free language, $\mathcal{B}(\mathcal{F})^{\mathbf{R}} = \{x^{\mathbf{R}} | x \in \mathcal{B}(\mathcal{F})\}$ is also context free, where $x^{\mathbf{R}}$ denotes the reversal of block $x$.
Thus a NPDA $M = \{Q,F, \Gamma, \delta, q_0, Z, E\}$ to recognize $\mathcal{B}(\mathcal{F})^{\mathbf{R}}$ exists.
Here a NPDA $M' = \{Q',\mathcal{A}, \Gamma', \delta', q'_0, Z', E'\}$ is defined as follows:
let $N = \{0,1,\cdots, p\}$ and
\begin{ul}{0.2}
\item $Q' = (Q \times \mathcal{A} \times N) \cup \{q'_0\}$,
\item $\Gamma' = \Gamma \cup \{Z'\}$,
\item $E' = \{(q, a, i) \in Q' | q \in E \land i = p\}$;
\end{ul}
next, $\delta'$ is determined by the followings:
\begin{ul}{0.2}
\item $\delta'(q_0', a, Z') = \{(\langle q_0, a, 0 \rangle ,Z)\}$ \quad for any $a \in \mathcal{A}$;
\item $\delta'(q_0', \epsilon, Z') = \{(\langle q_0, a, 1 \rangle ,Z) | ~ a \in \mathcal{A}\}$;
\item suppose that $q \in Q$, $a \in \mathcal{A}$, and $A \in \Gamma$;
if $b \in \mathcal{A}$ then 
\begin{equation}
\hspace*{-3em} \delta'(\langle q, a, 0 \rangle, b, A) = \{(\langle r,b,0 \rangle, g) | \exists f \in F ~ (r, g) \in \delta(q, f, A) \land f(b) = a\},
\end{equation}
else if $b = \epsilon$, then
\begin{eqnarray}
\delta'(\langle q, a, 0 \rangle, \epsilon, A) & = & \{(\langle r,a, 0 \rangle, g) | (r, g) \in \delta(q, \epsilon, A) \} \nonumber \\
 & & \cup ~ \{(\langle q, a, 1 \rangle, A)\};
\end{eqnarray}
\item for any $1 \leq i \leq p$,
\begin{eqnarray}
\hspace*{-5em} \delta'(\langle q, a, i \rangle, \epsilon, A) & = & \{(\langle r, b, i+1  \rangle, g) | b \in \mathcal{A}, \exists f \in F~ (r,g) \in \delta(q, f, A) \land f(b) = a \} \nonumber \\
 & & \cup ~  \{(\langle r,a, i \rangle, g) | (r, g) \in \delta(q, \epsilon, A) \}.
\end{eqnarray}
\end{ul}

Given an input $x \in \mathcal{A}^*$, $M'$ accepts $x$ iff there are $y = xa \in \mathcal{A}^{|x|+p}$ and $f \in F^{|x|+p-1}$ such that $M$ accepts $f$ and $y_{i-1} = f_{i}(y_{i})$ for $1 < i \leq |y|$.
Thus $y^{\mathbf{R}} \in \mathcal{D}_{\mathcal{F}}(p, x^{\mathbf{R}})$, so that $M'$ accepts $x$ iff $\mathcal{D}_{\mathcal{F}}(p, x^{\mathbf{R}}) \neq \emptyset$.
By lemma \ref{lemma:pumping_lemma}, there is a natural number $p$ such that $\mathcal{D}_{\mathcal{F}}(p, x^{\mathbf{R}}) \neq \emptyset$ iff $\lim_{n \rightarrow \infty}\mathcal{D}_{\mathcal{F}}(n, x^{\mathbf{R}}) \neq \emptyset$.
Hence there is a NPDA $M'$ such that $M'$ accepts $x$ iff $x^{\mathbf{R}} \in \mathcal{B}(X_{\mathcal{F}})$ by lemma \ref{lemma:condition_of_word_adscription}.
Accordingly, $\mathcal{B}(X_{\mathcal{F}})$ is a context free language.
\hfill \qed
\newline

\subsection{Context sensitive, recursive, and r.e.\ sets}
Let us consider the case in which the language of a functional shift $\mathcal{F}$ is not context free.
A class of dynamical systems, called generalized shifts, has been proposed by Moore \cite{Moore1991}, as corresponding to the class of languages more complicated than context-free languages.
The class of generalized shifts is equivalent to that of Turing machines, and they can be embedded in smooth maps in $\mathbb{R}^2$ or smooth flows in $\mathbb{R}^3$.

In this case, the problem of determining the predicate $x \in \mathcal{B}(X_{\mathcal{F}})$ is more difficult than in the above subsection.
This is because there does not necessarily exist a number $p$ to satisfy that for any $x \in \mathcal{A}^*$ $\mathcal{D}_{\mathcal{F}}(q, x) \neq \emptyset$ iff $\lim_{n \rightarrow \infty}\mathcal{D}_{\mathcal{F}}(n, x) \neq \emptyset$ in the case in which the language of a functional shift $\mathcal{F}$ is not context free, while it always exists in context free cases.

We suppose that $F$ is a set of bijections and $\mathcal{F}$ is a functional shift over $F$.
From the definition of $\mathcal{D}_{\mathcal{F}}$, it is clear that $\mathcal{D}_{\mathcal{F}}(0, x) \neq \emptyset$ is a sufficient condition for $\lim_{n \rightarrow \infty}\mathcal{D}_{\mathcal{F}}(n, x) \neq \emptyset$.
Then we can prove the following theorems.

\begin{theorem}\label{theorem:csg_to_csg_in_surjection_case}
{
\rm
Let $F$ be a set of bijections on $\mathcal{A}$, and $\mathcal{F}$ be a functional shift over $F$.
If $\mathcal{B}(\mathcal{F})$ is a context sensitive language, then $\mathcal{B}(X_{\mathcal{F}})$ is context sensitive.
}
\end{theorem}
\proof
Let $G = \{N, F, P, S\}$ be a context sensitive grammar to generate $\mathcal{B}(\mathcal{F})$.
$G' = \{N', \mathcal{A}, P', S'\}$ is determined by the following conditions:
\begin{ul}{0.2}
\item $N' = N \cup F \cup \{S'\}$, where $S' \not\in N \cup F$.
\item $P'$ contains only productions satisfying the following restrictions:
	\vspace{0.5ex}
	\begin{ul}{0.2}
	\item for any $a \in \mathcal{A}$, $S' \rightarrow aS \in P'$ and $S' \rightarrow a \in P'$;
	\item for any $\alpha,\beta \in (N \cup F)^*$, if $\alpha \rightarrow \beta \in P$ then $\alpha \rightarrow \beta \in P'$;
	\item if $f \in F$ and $a \in \mathcal{A}$, then $af \rightarrow af(a) \in P'$;
	\end{ul}
	\vspace{0.5ex}
\end{ul}
Then $G'$ is a context sensitive grammar because if $\alpha \rightarrow \beta \in P'$ then $|\alpha| \leq |\beta|$.

Let $x \in L(G')$.
Since there is a block $f \in L(G) = \mathcal{B}(\mathcal{F})$ such that $|f| = |x|-1$ and $x_{i+1} = f_{i}(x_i)$ for $1 \leq i < |x|$, then $\mathcal{D}_{\mathcal{F}}(0,x) \neq \emptyset$.
Thus $\lim_{n \rightarrow \infty}\mathcal{D}_{\mathcal{F}}(n, x) \neq \emptyset$ because every function in $F$ is a bijection, so that $x \in \mathcal{B}(X_{\mathcal{F}})$ by lemma \ref{lemma:condition_of_word_adscription}.
Hence $L(G') \subset \mathcal{B}(X_{\mathcal{F}})$.

Conversely, let $x \in \mathcal{B}(X_{\mathcal{F}})$.
Clearly there is a block $f \in \mathcal{B}_{|x|-1}(\mathcal{F})$ such that $x_{i+1} = f_{i}(x_i)$ for any $1 \leq i < |x|$.
Thus $f$ can be derived from $S$ in $G$.
Then
\begin{eqnarray}
S' & \underset{G'}{\Longrightarrow} & x_1S \nonumber \\
 & \overset{*}{\underset{G'}{\Longrightarrow}} & x_1f \nonumber \\
 & \overset{*}{\underset{G'}{\Longrightarrow}} & x,
\end{eqnarray}
so that $x \in L(G')$.
Hence $\mathcal{B}(X_{\mathcal{F}}) \subset L(G')$.
Accordingly $L(G') = \mathcal{B}(X_{\mathcal{F}})$ and $\mathcal{B}(X_{\mathcal{F}})$ is a context sensitive language.
\hfill \qed

\begin{theorem} \label{theorem:recursively_enumerable_in_surjection_case}
{
\rm
Let $F$ be a set of bijections on $\mathcal{A}$, and $\mathcal{F}$ be a functional shift over $F$.
If $\mathcal{B}(\mathcal{F})$ is r.e., then $\mathcal{B}(X_{\mathcal{F}})$ is r.e.
}
\end{theorem}
\proof
Suppose that $G$ is a type-0 grammar to generate $\mathcal{B}(\mathcal{F})$, and $G'$ is defined in the similar way to the proof of theorem \ref{theorem:csg_to_csg_in_surjection_case}.
As discussed in that proof, $G'$ is also a type-0 grammar and $L(G') = \mathcal{B}(X_{\mathcal{F}})$.
\hfill \qed
\newline

In the general case in which some functions in $F$ may be not bijections, it is difficult to determine where the languages of generated shifts are located in the Chomsky hierarchy.
To locate a collection of forbidden blocks is, however, an easier task than studying this problem.

\begin{theorem} \label{theorem:forbidden_blocks_is_csl}
{
\rm
Suppose that $\mathcal{F}$ is a functional shift and $\mathcal{B}(\mathcal{F})$ is context sensitive.
Then there is a context sensitive language $\mathtt{F}$ of forbidden blocks such that $X_{\mathtt{F}} = X_{\mathcal{F}}$.
}
\end{theorem}
\proof
Let
\begin{equation}
\mathtt{F} = \{x \in \mathcal{A}^* | \mathcal{D}_{\mathcal{F}}(0, x) = \emptyset\}.
\end{equation}
Now we show that $\mathtt{F}$ is a collection of forbidden blocks such that $X_{\mathtt{F}} = X_{\mathcal{F}}$.
Suppose that $x$ is a bi-infinite sequence in $\mathcal{A}^{\mathbb{Z}}$ such as $x \not\in X_{\mathcal{F}}$.
Then there is a block $y$, which is not contained in $\mathcal{B}(X_{\mathcal{F}})$, occurring in $x$ .
Thus, by lemma \ref{lemma:condition_of_word_adscription}, a natural number $n$ such as $\mathcal{D}_{\mathcal{F}}(n, y) = \emptyset$ exists.
For any $a \in \mathcal{A}^n$, $ay \in \mathtt{F}$ because $\mathcal{D}_{\mathcal{F}}(0, ay) = \emptyset$.
Since there is a block $a \in \mathcal{A}^n$ such that $x$ contains $ay$, some blocks in $\mathtt{F}$ occur in $x$.
Conversely, suppose that $x \in \mathcal{A}^{\mathbb{Z}}$ contains a block $y \in \mathtt{F}$.
Then it is clear that $x \not\in X_{\mathcal{F}}$.
Accordingly, $x \not\in X_{\mathcal{F}}$ iff there is a block in $\mathtt{F}$ which occurs in $x$.
Hence $X_{\mathtt{F}} = X_{\mathcal{F}}$.

\begin{figure}[hbtp]
\setlength{\unitlength}{0.7mm}
	\begin{center}
	\includegraphics{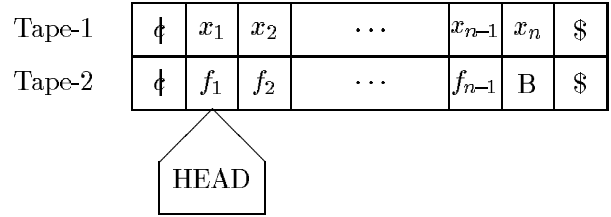}
	\end{center}
	\caption{Illustration of the linear bounded automaton $M_2$.}
	\label{figure:lba_illustration}
\end{figure}

Next, we will explain the existence of a linear bounded automaton (LBA) which can recognize $\mathtt{F}$.
Let $M_1$ be a LBA to compute the following function
\begin{equation} \label{equation:binding_function}
\varphi(x) = \left\{ \begin{array}{ll}
	1	& \qquad \mbox{if $x \in \mathcal{B}(\mathcal{F})$}, \\
	0	& \qquad \mbox{if $x \not\in \mathcal{B}(\mathcal{F})$}. \\
	\end{array} \right.
\end{equation}
We construct a LBA $M_2$ with two separate tapes as follows.
First, a block $x \in \mathcal{A}^*$ is inscribed on tape-1, and $f \in F^{|x|-1}$ on tape-2 (see Figure \ref{figure:lba_illustration}).
Then $M_2$ carries out the following operations.
\begin{ol}{1}
\item The machine $M_2$ begins with the head resting in anticipation on the left most cell.
The machine repeatedly moves right and reads the cell value beneath the head until the right most cell.
When the machine finds $i$ such that $x_{i+1} \neq f_i(x_i)$, $M_2$ accepts $\langle x,f \rangle$ and halts.
\item The machine $M_2$ calls the subroutine $M_1$ with the block $f$ on the tape-2, which returns the answer ``$0$'' or ``$1$'' as appropriate.
If the answer is ``$0$'', that is, $M_1$ does not accept $f$, then $M_2$ accepts $\langle x,f \rangle$.
In the other case, $M_2$ does not accept $\langle x,f \rangle$.
\end{ol}
Now we consider a machine $M$ such that, for any input $x$, if $M_2$ accepts $\langle x,f \rangle$ for all $f \in F^{|x|-1}$ then $M$ accepts $x$.
Since to construct a LBA to enumerate $F^{|x|-1}$ is an easy task, from this explicit definition we can get the LBA $M$.
For any $x$, $M$ accepts $x$ iff $\mathcal{D}_{\mathcal{F}}(0, x) = \emptyset$ because there is no $f \in \mathcal{B}_{|x|-1}(\mathcal{F})$ such that $x_{i+1} = f_i(x_i)$ for any $1 \leq i < |x|$.
Hence $M$ is a LBA to recognize $\mathtt{F}$.
\hfill \qed
\newline

Moreover, by using a proof similar to theorem \ref{theorem:forbidden_blocks_is_csl}, we can prove that if the language of a functional shift is a recursive set, then $\mathtt{F}$ is also recursive.
A language is a recursive set if there exists a Turing machine which recognizes the language and always halts.

\begin{theorem} \label{theorem:forbidden_blocks_is_recursively}
{
\rm
Suppose that $\mathcal{F}$ is a functional shift and $\mathcal{B}(\mathcal{F})$ is recursive.
Then there is a recursive set $\mathtt{F}$ of forbidden blocks such that $X_{\mathtt{F}} = X_{\mathcal{F}}$.
}
\end{theorem}
\proof
Since $\mathcal{B}(\mathcal{F})$ is a recursive set, there is a Turing machine $M_3$ to compute the function $\varphi$ in equation (\ref{equation:binding_function}).
Suppose that $M_2$ in the proof of theorem \ref{theorem:forbidden_blocks_is_csl} calls subroutine $M_3$ instead of $M_1$.
Then $M_2$ and $M$ are Turing machines that always halt after a finite amount of time, and $M$ recognizes $\mathtt{F}$.
Thus, $\mathtt{F}$ is a recursive set.
\hfill \qed

\subsection{The language of a generated shift beyond r.e.}
In this subsection, we prove that there is a functional shift $\mathcal{F}$ satisfying that $\mathcal{B}(\mathcal{F})$ is r.e.\ and $\mathcal{B}(X_{\mathcal{F}})$ is not r.e.
Here a set $A$ is r.e.\ iff the predicate $x \in A$ is partially decidable, i.e., the partial characteristic function
\begin{equation}
f(x) = \left\{ \begin{array}{ll}
	1	& \qquad \mbox{if $x \in A$},  \\
	\mbox{undefined}	& \qquad \mbox{if $x \not\in A$}, \\
	\end{array} \right.
\end{equation}
is computable.
Note that if a predicate $P(x)$ is partially decidable and undecidable, then $\neg P(x)$ is not partially decidable.
For example, the following predicate
\[
\mbox{`a Turing machine of the G\"{o}del number $x$ eventually stops on input $x$'}
\]
is partially decidable and undecidable.
In the proof of the next theorem, we show that $x \in \mathcal{B}(X_{\mathcal{F}})$ iff a Turing machine of the G\"{o}del number $x$ never halts on input $x$, in order to prove $x \in \mathcal{B}(X_{\mathcal{F}})$ is not partially decidable.

\begin{theorem} \label{theorem:language_of_generated_shift_beyond_r.e.}
{
\rm
There is a function shift $\mathcal{F}$ such that $\mathcal{B}(\mathcal{F})$ is r.e.\ and $\mathcal{B}(X_{\mathcal{F}})$ is not r.e.
}
\end{theorem}
\proof
For any $x \in \{01^n\delta| n \in \mathbb{N}\}$, let
\begin{equation}
num(x) = \mbox{the number of $1$'s occurring in $x$}
\end{equation}
and a Turing machine of the G\"{o}del number $n$ be denoted by $T_n$.
We will show a functional shift $\mathcal{F}$ such that $\mathcal{B}(\mathcal{F})$ is r.e.\ and
\begin{equation}
x \in \mathcal{B}(X_{\mathcal{F}}) \Leftrightarrow \mbox{$T_{num(x)}$ never halts on input $x$}.
\end{equation}

Let $\mathcal{A} = \{0,1,\delta\}$, $F = \{f, g, h\}$ be a set of maps defined by 
\begin{center}
\begin{tabular}{|cccc|}
\hline
$x$ & $f(x)$ & $g(x)$ & $h(x)$ \\
\hline
$0$ & $0$ & $1$ & $\delta$ \\
$1$ & $\delta$ & $1$ & $\delta$ \\
$\delta$ & $\delta$ & $\delta$ & $\delta$ \\
\hline
\end{tabular}
\end{center}
and $\mathcal{F}$ be a functional shift over $F$ such that $\mathcal{B}(\mathcal{F})$ is r.e.
Suppose that a Turing machine $M$ to recognize $\mathcal{B}(\mathcal{F})$ satisfies the following conditions:
\begin{ul}{0.2}
\item $T_{num(x)}$ does not halt on input $x$ before time $t$ iff $M$ accepts $f^{t}g^{num(x)}f$,
\item $T_{num(x)}$ halts on input $x$ at time $t$ iff $M$ accepts $hf^tg^{num(x)}f$,
\end{ul}
where $x \in \{01^n\delta| n \in \mathbb{N}\}$.
Since it is clear that $num(x)$ and the emulation of $T_{num(x)}$ are in fact computable functions, $M$ exists, by Church's thesis.

For any $x \in \{01^n\delta| n \in \mathbb{N}\}$, $\mathcal{D}_{\mathcal{F}}(t, x) \neq \emptyset$ iff $T_{num(x)}$ does not halt on input $x$ before time $t+1$.
Accordingly, 
\begin{eqnarray}
\mbox{$T_{num(x)}$ never halts on input $x$} & \Leftrightarrow & \lim_{t \rightarrow \infty}\mathcal{D}_{\mathcal{F}}(t, x) \neq \emptyset \nonumber \\
& \Leftrightarrow & x \in \mathcal{B}(X_{\mathcal{F}}),
\end{eqnarray}
by lemma \ref{lemma:condition_of_word_adscription}.
Hence the predicate $x \in \mathcal{B}(X_{\mathcal{F}})$ is not partially decidable, so that $\mathcal{B}(X_{\mathcal{F}})$ is not r.e.
\hfill \qed

\section{Discussion} \label{section:discussion}
\begin{figure}[hbtp]
\setlength{\unitlength}{0.7mm}
	\begin{center}
	\includegraphics{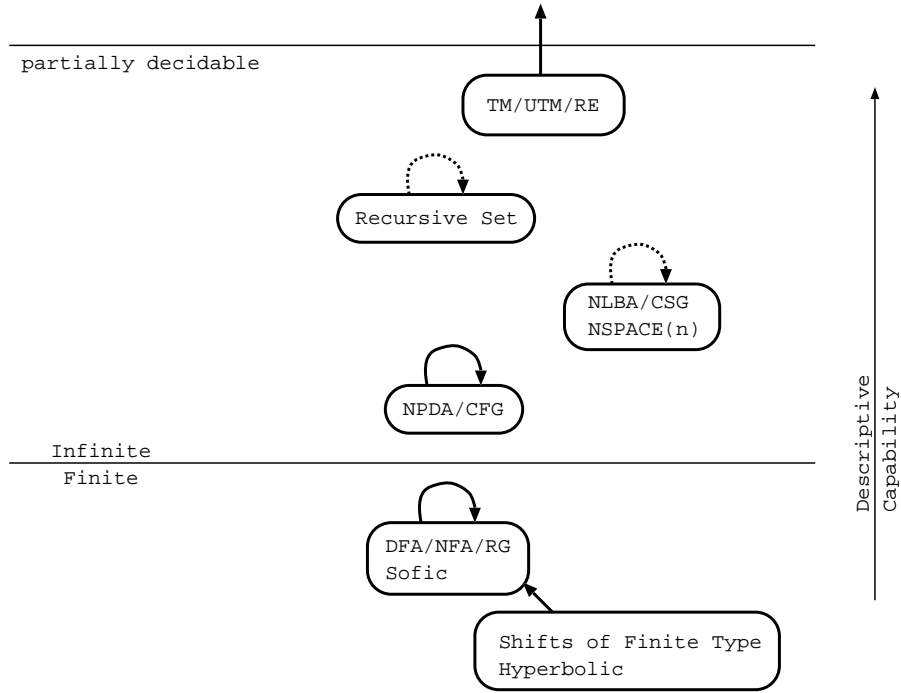}
	\end{center}
	\caption{The computational hierarchy of the languages of functional shifts and generated shifts.
	Every class of the languages of functional shifts is the same as or smaller than that of generated shifts.
	For functional shifts with r.e.\ language, we have generated shifts whose language is beyond r.e.
	The solid arrow from $A$ to $B$ denotes that $A$ is a class of the languages of functional shifts and $B$ is that of shift spaces generated by the functional shifts.
	The broken arrow from $A$ to $B$ denotes that $A$ is the same as the case of solid arrow and $B$ is a class of sets of forbidden blocks which can describe generated shifts given by the functional shifts.
	D = deterministic, N = nondeterministic, U = universal, G = grammar, A = automata, TM = Turing machine, RE = recursively enumerable, LBA = linear bounded A, PDA = pushdown A, FA = Finite A, CSG = context sensitive G, CFG = context free G, RG = regular G.}
	\label{figure:descriptive_capability}
\end{figure}

To study the relationship between dynamics and meta-dynamics, we have compared functional shifts with generated shifts, in both dynamical systems and computational terms.
In section \ref{section:entropy}, we have proved that the entropy of a functional shift gives the upper limit for that of a generated shift given by the functional shift.
In other words, every functional shift is at least as complex as the generated shift from the standpoint of dynamics.
In section \ref{section:computation}, considering the language of a shift space as a formal language, we have shown that there is a case in which the language of a functional shift is simpler than that of a shift space generated by the functional shift.
Figure \ref{figure:descriptive_capability} shows the summary of results, in which the languages of functional shifts and generated shifts are located in the Chomsky hierarchy.
It shows clearly that the class of the languages of functional shifts is the same as or smaller than that of the generated shifts given by them.
From those results, we may consider that the complexity of dynamics does not correspond to that of computation under the generative operation introduced here.
Moreover, the viewpoints of both dynamics and computation give us opposite results concerning the complexity of systems.
This means that an analysis of the complexity of systems surely depends on how we select a measure of complexity.
Thus it is important to study dynamical systems from several viewpoints, for example, those of both dynamics and computation, when we analyze the complexity of the systems.

It is interesting that the class of languages of functional shifts is equal to that of generated shifts, in the case in which the languages of the functional shifts are regular or context free, while there is no equivalence in the r.e.\ cases.
The cause is conjectured to be that the equivalency depends on whether there exists a natural number $n$ such that $\mathcal{D}_{\mathcal{F}}(n,x) \neq \emptyset$ iff $x \in \mathcal{B}(X_{\mathcal{F}})$ for any $x \in \mathcal{A}^*$.
Lemma \ref{lemma:pumping_lemma} and theorem \ref{theorem:cfg_to_cfg} confirm our presumption.
Theorem \ref{theorem:csg_to_csg_in_surjection_case} and theorem \ref{theorem:recursively_enumerable_in_surjection_case} also support it, because if any functions are bijections then $\mathcal{D}_{\mathcal{F}}(0,x) \neq \emptyset$ iff $x \in \mathcal{B}(X_{\mathcal{F}})$.

Let us discuss the existence of systems in which some predicates of dynamical systems theory are not partially decidable.
In theorem \ref{theorem:language_of_generated_shift_beyond_r.e.}, we have proved that there is a functional shift $\mathcal{F}$ satisfying that the language of $\mathcal{F}$ is r.e.\ and that of $X_{\mathcal{F}}$ is not r.e.
Note that the predicate $x \in \mathcal{B}(X)$ means that the subset $\{y \in \mathcal{A}^{\mathbb{Z}} | ~ y_{[1,|x|]} = x\}$ of $\mathcal{A}^{\mathbb{Z}}$ contains part of an invariant set $X$.
Many problems about dynamical systems can be resolved into this predicate.
For example, we can consider such problems as follows:
\begin{ol}{1}
\item Given open sets $Y,Z \subset X$, is there a point $z \in Y$ that falls into $Z$ under the shift map on $X$?
\item Is the shift map on $X$ topologically transitive?
\end{ol}
Problem (1) is considered to be a prediction problem of orbits in a dynamics system.
The reason why problem (1) resolves itself into the predicate $x \in \mathcal{B}(X)$ is that if $u, v \in \mathcal{B}(X)$, $Y = \{y \in X | ~ y_{[1,|u|]} = u\}$ and $Z = \{y \in X | ~ y_{[1,|v|]} = v\}$, then there is a block $w$ such that $uwv \in \mathcal{B}(X)$ iff (1) is true.
Since problem (2) can be reduced to (1), problem (2) contains the predicate as a subproblem.
Thus, a shift space $X_{\mathcal{F}}$ is so complex that those predicates are not necessarily partially decidable, even if such predicates of $\mathcal{F}$ are partially decidable.
As a dynamical system in which those problems are not partially decidable, we may consider the system with riddled basins \cite{Alexander1992,Ott2002}.
In fact, probably no algorithm exists which is able to assess problem (1) in a finite number of steps if Z is a riddled basin \cite{Badii1997,Blum1993}.

Generally, the automaton to recognize a set not to be r.e.\ is called a super-Turing machine \cite{Stannett1990,Copeland1998}.
Every super-Turing machine recognizes a set to be beyond r.e.\ by using infiniteness, for example the property of the real number, which usual Turing machines do not have.
Since the languages of shift spaces generated by functional shifts with r.e.\ languages are not r.e., those shift spaces have relevance to super-Turing machines.
Some classes of sets to be beyond r.e.\ have been discussed in the field of analog computation \cite{Siegelmann1995,Siegelmann1998,Siegelmann1999,Ben-Hur2002,Ben-Hur2003,Hamkins2000}.
Hamkins and Lewis, for instance, analyze classes of computations with infinitely many steps, and investigate computability and decidability on the reals \cite{Hamkins2000}.
Notice that shift spaces are usually continuums, because every shift space is defined as a collection of bi-infinite sequences.
Hence, constructing a generated shift from a functional shift is an operation on a continuum, including tasks such as infinite mapping $x_{i+1} = f_i(x_i)$ for all $i \in \mathbb{Z}$.
This operation, which is implicitly involved in the definition of functional shifts and generated shifts, often affects properties of the shifts themselves.
For example, by using such an operation, we show that the predicate $x \in \mathcal{B}(X_{\mathcal{F}})$ is not partially decidable in the proof of theorem \ref{theorem:language_of_generated_shift_beyond_r.e.}.
Therefore, our framework could be related to super-Turing machines and analog computation.
Further analysis from the viewpoints of them is a future research topic.

We have introduced the framework of functional shifts as a model of dynamic change of functions.
Since we can consider that a bi-infinite sequence of function $(f_i)_{i\in \mathbb{Z}}$ denotes the evolution of maps, functional shifts represent dynamics of functions.
Generated shifts also represent the dynamics determined by $x_{i+1} = f_i(x_i)$.
Considering the shift map on $X_{\mathcal{F}}$ as dynamics, we can regard that on $\mathcal{F}$ as meta-dynamics.
Let us focus on the operation to generate shift spaces from functional shifts, i.e., to construct dynamics from meta-dynamics.
By theorem \ref{theorem:language_of_generated_shift_beyond_r.e.}, such operation includes a task to generate sets not to be r.e.\ in spite of the fact that any functions in $F$ are computable.
Thus, as we discussed, when we study dynamical systems modified by meta-dynamics, the existence of complex dynamics should be taken into account.

Finally, we consider self-modifying systems, in which rules governing a system are used to change the rules themselves.
We have difficulty in achieving a direct representation of the dynamics of a self-modifying type.
The cause of the difficulty is that functions and states cannot be perfectly separated in self-modifying systems.
Research into the function dynamics of the self-modifying type has recently become a subject of special interest in the study of these systems.
Functional dynamics is an example of a self-modifying system, because the change of function $f$ is determined by a self-reference term $f \circ f$ \cite{Kataoka2000,Kataoka2001,Kataoka2003}.
As another example, objects in algorithmic chemistry encode functions which change the objects themselves \cite{Fontana1992,Fontana1994}.
To describe self-modifying systems, we must take the self-referential nature of a dynamical system into account.
We can represent this characteristic using functional shifts as follows \cite{Namikawa2002}.
Let us consider that a conjugacy from a functional shift to a generated shift given by the functional shift is a `self-reference code' between functions and states.
Considering that each sequence of functions is equal to a sequence of symbols corresponding to itself under the code, we can regard the functional shift as self-modifying.
For instance, the functional shift $\mathcal{F}$ given in example \ref{example:full_shift} or \ref{example:sturmian_shift} has conjugacy to $X_{\mathcal{F}}$, so that the functional shifts in these examples are regarded as self-modifying systems.
However, the relationship between the systems described by functional shifts and other systems introduced by \cite{Kataoka2000,Kataoka2001,Kataoka2003,Fontana1992,Fontana1994} is not clear.

\section{Conclusion} \label{section:conclusion}
We have investigated shift dynamics called functional shifts within both dynamics and computational frameworks.
From the dynamical viewpoint, we have proved that the entropy of a functional shift is not less than that of a shift space generated by the functional shift.
This means that functional shifts generate less complex shift spaces than themselves.
On the other hand, we have compared functional shifts with generated shifts in terms of the Chomsky hierarchy (see Figure \ref{figure:descriptive_capability}).
We have proved that any class of the languages of shift spaces is at least as large as that of the functional shifts that generate the shift spaces.
Furthermore, we have shown that there is a class of the languages of functional shifts, which is strictly smaller than that of generated shifts given by the functional shifts.
From those results, we have argued that the viewpoints of dynamics and computation give us opposite results about complexity of systems.

We have shown a new class of shift spaces, generated shifts whose languages are not r.e.\ if the languages of functional shifts to give the generated shifts are r.e.
The shift map over some shift spaces in the class has very unpredictable dynamics.
This new class gives us a way to study dynamical systems from the viewpoint of analog computation.

\section*{Acknowledgment} 
The authors wish to thank Yuzuru Sato, Ichiro Tsuda, Hiroakira Ono, and Naoto Kataoka for helpful discussions and comments.
We are grateful to the anonymous reviewers for their critical reading of the manuscript and significant comments.
We would like to thank Judith Anne Steeh for her assistance in English editing.
This work is partly supported by Research Fellowship Program of Canon Foundation in Europe, and a Grant-in-Aid for Scientific Research (No.12780269) from the Ministry of Education, Culture, Sports, Science and Technology of Japan and by the Japan Society for the Promotion of Science.

\appendix
\section{Review of the Chomsky hierarchy} \label{section:appendix_a}
The Chomsky hierarchy is a containment hierarchy of classes of formal grammars that generate formal languages.
This hierarchy was described by Noam Chomsky\cite{Chomsky1959}.

A \textit{formal grammar} (or \textit{type-0 grammar}) is a 4-tuple $(V_N,V_T,P,S)$, where
\begin{ul}{0.2}
\item $V_N$ is a finite set of \textit{nonterminal symbols};
\item $V_T$ is a finite set of \textit{terminal symbols} such as $V_N \cap V_T = \emptyset$, and $V = V_N \cup V_T$ is called the set of \textit{grammar symbols};
\item $P$ is a finite collection of \textit{productions} which are of the form $\alpha \rightarrow \beta$ with $\alpha \in V^{+}$ and $\beta \in V^{*}$,
\item $S \in V_N$ is a designated symbol called the \textit{start symbol}.
\end{ul}
Given a formal grammar $G = (V_N,V_T,P,S)$, the \textit{derivation relation} $\underset{G}{\Longrightarrow} \subset V^* \times V^*$ is defined by
\begin{equation}
\gamma \alpha \delta \underset{G}{\Longrightarrow} \gamma \beta \delta \quad \mbox{iff} \quad \alpha \rightarrow \beta \in P,
\end{equation}
where $\alpha,\beta,\gamma,\delta \in V^*$.
The transitive closure of $\underset{G}{\Longrightarrow}$ is denoted by $\overset{+}{\underset{G}{\Longrightarrow}}$, and the reflexive and transitive closure of $\underset{G}{\Longrightarrow}$ is denoted by $\overset{*}{\underset{G}{\Longrightarrow}}$.
The language generated by $G$ is the set 
\begin{equation}
L(G) = \{w \in V_T^{*} | S \overset{*}{\underset{G}{\Longrightarrow}} w\}.
\end{equation}
A language $L \subset V_T^*$ is a \textit{formal language} (or \textit{type-0 language}) iff $L = L(G)$ for some formal grammar $G$.

The Chomsky hierarchy consists of classes of regular grammars, context free grammars, context sensitive grammars, and type-0 grammars.
Regular, context free, and context sensitive grammars are more restrictive than formal grammars.
\begin{ul}{0.2}
\item A regular grammar (type-3 grammar) is a formal grammar $G = (V_N,V_T,P,S)$, such that the productions are of the form $\alpha \rightarrow \beta$ with $\alpha \in V_N$ and $\beta \in V_T \cup V_T \times V_N$.
A language $L \subset V_T^*$ is a regular language iff $L = L(G)$ for some regular grammar $G$.
\item A context free grammar (type-2 grammar) is a formal grammar $G = (V_N,V_T,P,S)$, such that the productions are of the form $\alpha \rightarrow \beta$ with $\alpha \in V_N$ and $\beta \in V^{+}$.
A language $L \subset V_T^*$ is a context free language iff $L = L(G)$ for some context free grammar $G$.
\item A context sensitive grammar (type-1 grammar) is a formal grammar $G = (V_N,V_T,P,S)$, such that the productions are of the form $\alpha \rightarrow \beta$ satisfying $|\alpha| \leq |\beta|$.
A language $L \subset V_T^*$ is a context sensitive language iff $L = L(G)$ for some context sensitive grammar $G$.
\end{ul}
From the above definition, every regular grammar is context free, every context free grammar is context sensitive and every context sensitive grammar is type-0.
Moreover, these are all proper inclusions.

It is known that there exists automaton corresponding to each formal language belonging to the Chomsky hierarchy.
Every type-0 language can be recognized by a Turing machine, where a Turing machine is a finite state machine moving left and right on a tape, on which a string of symbols in some finite alphabet is written.
The language recognized by a Turing machine is defined as all the strings on which it halts.
These languages are also known as the recursively enumerable (r.e.) languages.
Every context sensitive language can be recognized by a linear bounded automaton which is a nondeterministic Turing machine whose tape is bounded by the length of the input string.
Every context free language can be recognized by a nondeterministic pushdown automaton which is a finite state automaton having a stack.
Every regular language can be recognized by a finite state automaton which has no memory.
Thus, the complexity from the Chomsky hierarchy is based on the memory size of automata to recognize languages.
The table 1 summarizes each of the Chomsky's four types of grammars, the class of languages each grammar generates, and the type of automaton that recognizes each language.

\begin{table}[hbtp]

\begin{center}
{\small \textbf{Table 1} Summarize each of the Chomsky's four types of grammars, languages, and automata.}

\vspace{0.2cm}

\begin{tabular}{|c|c|c|}
\hline
Grammar & Language & Automaton \\
\hline
Type-0 & Recursively enumerable & Turing machine \\
\hline
Type-1 & Context sensitive & Linear bounded automaton \\
\hline
Type-2 & Context free & Nondeterministic pushdown automaton \\
\hline
Type-3 & Regular & Finite state automaton \\
\hline
\end{tabular}
\end{center}
\end{table}

\end{document}